\def
\papertitle{Interpretable timbre synthesis using variational autoencoders regularized on timbre descriptors}
\def\paperauthorA{Anastasia Natsiou}
\def\paperauthorB{Luca Longo}
\def\paperauthorC{Se\'{a}n O'Leary}

\documentclass[twoside,a4paper]{article}
\usepackage{etoolbox}
\usepackage{dafx_23}
\usepackage{amsmath,amssymb,amsfonts,amsthm}
\usepackage{euscript}
\usepackage[T1]{fontenc}
\usepackage[utf8]{inputenc}
\usepackage{ifpdf}
\usepackage[english]{babel}
\usepackage{caption}
\usepackage{subfig} 
\usepackage{color}

\input glyphtounicode
\ninept

\newcounter{numauth}
\newcounter{listcnt}
\newcommand\authcnt[1]{\ifdefined#1 \stepcounter{numauth} \fi}

\newcommand\addauth[1]{
\ifdefined#1 
\stepcounter{listcnt}
\ifnum \value{listcnt}<\value{numauth}
\appto\authorslist{, #1}
\else
\appto\authorslist{~and~#1}
\fi
\fi}
\authcnt{\paperauthorB}
\authcnt{\paperauthorC}
\authcnt{\paperauthorD}
\authcnt{\paperauthorE}
\authcnt{\paperauthorF}
\authcnt{\paperauthorG}
\authcnt{\paperauthorH}
\authcnt{\paperauthorI}
\authcnt{\paperauthorJ}
\def\authorslist{\paperauthorA}
\addauth{\paperauthorB}
\addauth{\paperauthorC}
\addauth{\paperauthorD}
\addauth{\paperauthorE}
\addauth{\paperauthorF}
\addauth{\paperauthorG}
\addauth{\paperauthorH}
\addauth{\paperauthorI}
\addauth{\paperauthorJ}

\usepackage{times}

\newif\ifpdf
\ifx\pdfoutput\relax
\else
   \ifcase\pdfoutput
      \pdffalse
   \else
      \pdftrue
\fi

\ifpdf 
  \usepackage[pdftex,
    pdftitle={\papertitle},
    pdfauthor={\authorslist},
    pdfsubject={Proceedings of the 26th International Conference on Digital Audio Effects (DAFx23)},
    colorlinks=false, 
    bookmarksnumbered, 
    pdfstartview=XYZ 
  ]{hyperref}
  \usepackage[pdftex]{graphicx}
\else 
  \usepackage[dvips]{epsfig,graphicx}
  \usepackage[dvips,
    pdftitle={\papertitle},
    pdfauthor={\authorslist},
    pdfsubject={Proceedings of the 26th International Conference on Digital Audio Effects (DAFx23)},
    colorlinks=false, 
    bookmarksnumbered, 
    pdfstartview=XYZ 
  ]{hyperref}
\fi
\usepackage[hypcap=true]{caption}
\title{\papertitle}

\affiliation{
\paperauthorA, \ \paperauthorB, \ and \paperauthorC \thanks{\vspace{-3mm}}}
{ Technological University Dublin\\ Dublin, Ireland \\
{\{\tt \href{mailto:anastasia.natsiou@tudublin.ie}{anastasia.natsiou}, \tt \href{mailto:luca.longo@tudublin.ie}{luca.longo}, \tt \href{mailto:sean.oleary@tudublin.ie}{sean.oleary}}\} @tudublin.ie
}

\begin{document}
\ifpdf 
  \DeclareGraphicsExtensions{.png,.jpg,.pdf}
\else  
  \DeclareGraphicsExtensions{.eps}
\fi


\maketitle
\begin{abstract}

Controllable timbre synthesis has been a subject of research for several decades, and deep neural networks have been the most successful in this area. Deep generative models such as Variational Autoencoders (VAEs) have the ability to generate a high-level representation of audio while providing a structured latent space. Despite their advantages, the interpretability of these latent spaces in terms of human perception is often limited. To address this limitation and enhance the control over timbre generation, we propose a regularized VAE-based latent space that incorporates timbre descriptors. Moreover, we suggest a more concise representation of sound by utilizing its harmonic content, in order to minimize the dimensionality of the latent space.

\end{abstract}

\section{Introduction}
\label{sec:intro}

The emergence of deep generative models has contributed to the development of natural and expressive music synthesizers \cite{engelddsp, engelgansynth}. One type of generative model, Variational Autoencoders (VAEs), can create a compact representation based on the distribution of the input data and this representation forms a \textit{latent space} \cite{kingma2013auto}. In this space, samples that are similar to each other are positioned closer. However, this proximity may not always align with human perception of similarity. Explainable artificial intelligence (XAI) has been employed in the past to regularize the latent space of generative models \cite{bryan-kinns_exploring_nodate, natsiou2023}. A regularized space can offer interpretability of the latent space and control of the synthesis process to generate sound with desired characteristics.

In the context of music, the perception of different instruments can be characterized by a multidimensional space called \textit{timbre space}. Timbre space is created by asking listeners to rate the dissimilarity between different instruments \cite{grey_multidimensional_1977, terasawa_perceptual_2005}. In order to utilize this information effectively, a recent approach in sound synthesis has been proposed to incorporate timbre space into the regularization of VAEs \cite{esling2018bridging}. This enables the model to generate sounds based on specific instrument timbres. While utilizing a timbre space has various advantages for generative models, creating a timbre space is not always feasible. The addition of a new instrument would necessitate new listening tests and analyses.

\textit{Timbre descriptors} are mathematical or statistical functions designed to capture various aspects of human perception of sound. They were originally designed by studying the listening tests and using the information obtained to associate mathematical formulas with timbre space. Several research studies have found that spectral centroid and attack time are the key audio features that are crucial for describing the timbre of musical instruments \cite{peeters_timbre_2011}. Our goal in this study is to utilize the timbre descriptors, particularly the spectral centroid and attack time, to construct a latent space that aligns with human perception. To maximize the compression of the latent space, we developed a novel input representation with lower dimensionality than spectrograms that focuses on the harmonic content of the instruments. The rest of the manuscript is divided as follows. Section 2 provides a literature review on the regularization of VAEs for sound synthesis. Details on the proposed representation along with VAEs and timbre descriptors are described in Section 3. Finally, Section 4 provides information on the experimental setup and Section 5 demonstrates the results.


\section{Related Work}
\label{sec:relatedwork}

Recent research has demonstrated the capability of unsupervised models to acquire invertible audio representations through the use of autoencoders \cite{engel2017neural}. However, autoencoders have certain limitations that prevent them from creating a coherent and comprehensible latent space, which, in turn, limits their ability to produce audio with specific characteristics. Variational autoencoders (VAEs) overcome this limitation by incorporating a Gaussian distribution into the latent space, which encourages local smoothness and provides interpretability of the latent variables \cite{tatar_latent_2021}. In numerous applications, however, this approach may not be sufficient.

To achieve additional disentanglement of the latent space, various approaches have been proposed. Autocoder \cite{franzson_autocoder_nodate} is a simplified VAE trained on audio samples with specific attributes. The main objective of the Autocoder is to produce outputs that fit within the distribution of the training data. Luo et al. \cite{luo_learning_2019} proposed a network with two encoders and one decoder to address the disentanglement of pitch and timbre in audio signals. The first encoder learns the pitch and the second learns the timbre, while the decoder reconstructs the original audio signal from the concatenation of the learned pitch and timbre representations. The network can be conditioned on the two categorical variables separately because pitch and timbre are independent of each other. Both of these studies used mel-spectrograms as the input data representation for their models.

An alternative method for creating a latent space that focuses on specific attributes was deployed in \cite{yang_deep_2019}. Instead of relying solely on the original input data, this method incorporates an additional representation for chords to help shape the latent space. The model was trained on 32 one-hot vectors of MIDI pitches that represented the rhythm for an analogy-making task. The VAE used Gated Recurrent Units (GRUs) conditioned on chromagrams that represented chords. The attributes of the latent space were implicitly formed by incorporating information about melody and chords in the loss function. However, other models took a different approach and explicitly designed the latent space by enforcing specific attributes to represent distinct characteristics of the sound. In \cite{bryan-kinns_exploring_nodate} the latent space consists of 256 dimensions, where the first four dimensions are specifically associated with rhythmic complexity, note range, note density, and average interval jump. The regularization is incorporated into the VAE as a loss function to the training objective. To achieve this, a musical metric value is computed for each item in a mini-batch for each attribute. The distance between each item's metric value and all other items' metric values is then calculated, resulting in a distance matrix. The distance matrix is used for the computation of the mean square error.

Our method follows an approach that is heavily influenced by \cite{esling2018bridging}. Esling et al. introduced an additional regularization loss to form a latent space based on human perception. The latent attributes were generated to satisfy the perceptual similarity ratings of listeners. The architecture of their network consists of a VAE that includes three dense layers and produces a latent space of 64 dimensions. The neural network architecture successfully captured a continuous and generalized timbre representation for a wide range of musical instruments. In our work, we aim for a regularized latent space that aligns with human perception using audio descriptors that are capable to represent timbre.

\section{Proposed Framework}
\label{sec:proposedFramework}
\subsection{Variational Autoencoders}

Variational Autoencoders (VAEs) \cite{kingma2013auto} are deep generative models with the ability to map high-dimensional observed data $x \in \mathbb{R}^d$ (such as audio samples or spectrograms) to a latent space $z \in \mathbb{R}^e$ with $d>e$. VAEs consist of an encoder and a decoder network. The latent variable is created by the encoder that uses a distribution $q_\theta(z|x)$ to approximate $p_\theta(z|x)$. The decoder then attempts to reconstruct the input data by approximating the distribution $p_\theta(x|z)$. The encoder and decoder are trained together to parametrize $_\theta$. To elaborate further, the encoder generates the mean $\mu_M$ and the covariance $\sigma_M$ to represent the Gaussian distribution function $N(z;\mu_M,\sigma_M^2I)$ in a latent space of M dimensions. The main goal of the network is to maximize the evidence lower bound by minimizing the Kullback-Leibler (KL) divergence between the distribution function $q(z|x)$ and the prior distribution function $p(z)$. The loss function of VAEs is represented in Eq. \ref{elbo}.

\begin{equation}
\label{elbo}
    \mathbb{E}[\log p(x|z)] - KL(q(z|x) \parallel p(z)) \leq \log p(x)
\end{equation}

The initial component of the equation evaluates the degree of similarity between the original data and the reconstructed data. The second component assesses the dissimilarity between the approximated posterior distribution $q(z|x)$ and the prior distribution $p(z)$. By including the KL divergence term, the learned posterior distribution is pushed towards the prior distribution, thereby promoting the regularization of the learned latent space representation.

\subsection{Proposed Representation}

In this study, we introduce a novel audio representation for capturing monophonic and harmonic sounds of musical instruments based on acoustic features. The aim of this approach is to generate a concise representation that can enhance the efficacy of deep neural networks. In order to achieve this goal, we are based on the observation that monophonic, and harmonic sounds can be represented by their fundamental frequency, the first seven harmonics, and the energy of the higher bands. A method for synthesizing from a representation like the one described can be accomplished by employing a sinusoidal model, similar to the approach described in the work of \cite{natsiou2022}.

In this work, we estimate the fundamental frequency using a pre-trained model of CREPE \cite{kim2018crepe}, and then we calculate the logarithm of the amplitudes of the first seven harmonics. The harmonics can be estimated as the integer multiples of the fundamental frequency:

\begin{equation}
    f_{n}=nf_{0}
    \label{freqs}
\end{equation}
where the variable $n$ denotes the specific harmonic number for each frame $i$ of the sound, where $n \in [1, 7]$. The first seven harmonics are selected as they are deemed to contain the most perceptually significant aspects of a note, thereby providing essential information on the spectral shape of the acoustic signal. The remaining spectral information is represented by the energy of the higher bands. The higher spectral content is divided into 4 bands using the Equal Rectangular Bandwidth (ERB) \cite{moore1983suggested}.








\subsection{Timbre Descriptors}
Timbre constitutes the quality of sound that is conceptually separated from pitch or loudness. It refers to the unique hearing "color" provided by each instrument or voice. Timbre descriptors are used to quantify and describe the timbre of a sound. The association between timbre descriptors and the perception of sound is strong. Certain timbre descriptors are correlated with specific percepts of sound, such as brightness, dullness, spectral tilt, and tonality \cite{peeters_timbre_2011}. However, many studies have shown that timbre can accurately be described only with spectral centroid and attack time \cite{handel_timbre_1995}.

\subsubsection{Spectral centroid}

The spectral centroid is a measure of the center of mass of the spectrum of a sound. It is calculated by weighting each frequency in the spectrum by its magnitude and averaging the result as it is demonstrated in Eq. \ref{centroidEq}.

\begin{equation}
centroid = \frac{\sum_{k = b_1}^{b_2}f_k M(f_k)}{\sum_{k = b_1}^{b_2}M(f_k)}
\label{centroidEq}
\end{equation}
where $b_1$ and $b_2$ are the band edges, $f_k$ is the frequency in the bin $k$, and M(f) is the magnitude of the frequency in the spectrum. The spectral centroid is a useful measure of the timbre of a sound, as it provides information about the distribution of energy across different frequencies in the sound. It is often used to describe the brightness or dullness of a sound \cite{peeters2004large}. Sounds with a high spectral centroid tend to be brighter and more focused, while sounds with a low spectral centroid tend to be duller and more diffuse.

\subsubsection{Attack time}
Attack time is a parameter that determines how quickly the amplitude of a sound increases from zero to its maximum level. Attack time is an important factor in shaping the envelope of a sound and has a significant impact on its timbre \cite{gordon_perceptual_1987}. The computation of attack time has been under investigation for many years. The most common methods include a fixed threshold. Thresholds can vary and effective ones can be from 10\% to 90\% of the maximum value of the energy envelope \cite{virtanen_acoustics_2018} or from 20\% to 80\% of the maximum value of the envelope \cite{klapuri2007signal}. In this work, we measure attack time as a fixed threshold between 10\% to 90\%.

\section{Experimental setup}
\label{sec:experimentalSetup}
\subsection{Dataset}

The NSynth dataset \footnote{\url{https://magenta.tensorflow.org/datasets/nsynth}}, containing a wide range of monophonic notes from various instruments, was utilized in the experiments. The dataset was filtered to include only samples that were harmonic, without variations in fundamental frequency or amplitude. The resulting subsample was composed of 101,911 training samples and 1,324 testing samples of various instruments, including guitar, bass, brass, keyboard, flute, organ, reed, and string, with a pitch range of 80Hz-2100Hz. The audio samples were preprocessed to create a representation that comprises the fundamental frequency, the logarithm of the amplitude of the first 7 harmonics, and 4 ERBs in overlapping segments of audio signals with a window of 690 and a hop size of 172.

\subsection{Hyperparameters}

The encoder is composed of two 2D convolutional layers with 32 filters, a kernel size of 3, a stride of 2, and the same padding. Two dense layers calculate the mean and variance of the Gaussian distribution. Sampling from the distribution creates a latent space with 14 dimensions. The decoder follows a mirrored architecture of the encoder generating the proposed representation. The ReLU is used as an activation function for the convolutional layers while the softmax function is applied to the output layer to form the generated normalized representation. The network is trained using the ADAM optimizer with an initial learning rate of 0.001 in batches of size 128. The loss function includes a binary cross-entropy reconstruction loss, a regularization loss consisting of a KL-divergence term, and a mean absolute error of the spectral centroid and attack time. The model was trained using the TensorFlow library\footnote{\url{https://www.tensorflow.org/}} on a Tesla P100 GPU.

\subsection{Evaluation}
For the evaluation of the reconstruction capacity of the VAE, we used the Mean Squared Error (MSE) and the Structural SIMilarity (SSIM) index between the original and generated audio representation based on the harmonic content. The latent space was visualized by projecting the high-dimensional features into a two-dimensional space. For the dimensionality reduction, we used the Stochastic Neighbor Embedding (t-SNE) algorithm \cite{van2008visualizing} with PCA initialization and perplexity of 50.

\section{Results and Conclusion}
\label{sec:results}

\begin{figure}[ht]
\centerline{\includegraphics[scale=0.33]{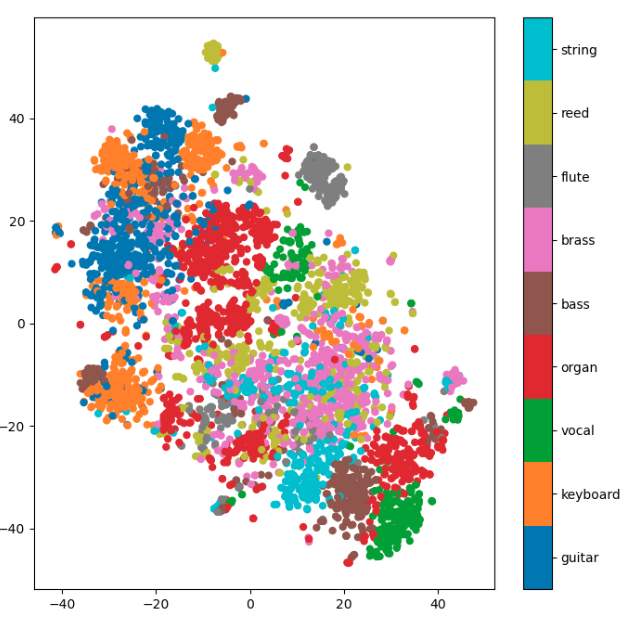}}
\caption{\label{latentplot}{\it 2D projection of the regularized latent space}}
\end{figure}

Table \ref{tab:rec} presents the reconstruction error for the model with and without the regularization of the timbre descriptors. The results showed that convolutional VAEs are able to adequately reconstruct the proposed representation. However, the additional regularization decreased the quality of the generated samples but it provided a clear representation of the latent space. Fig. \ref{latentplot} displays the 14-dimensional regularized latent space projected into a 2-dimensional space. It is evident from the figure that different instrument categories occupy distinct positions in the space, indicating that the model is capable of distinguishing between instrument timbre. However, some instrument categories may be broad and have multiple variations, resulting in clusters in different regions of the space. The obtained results demonstrate that using timbre descriptors to regularize VAEs can lead to a high-level latent representation that is interpretable. This research work illustrates an early investigation of the merging of timbre descriptors with deep generative models.

\begin{table}[ht]
  \caption{\itshape Reconstruction error}
	\centering
	\begin{tabular}{|c|c|c|}
		\hline
		  Model & MSE & SSIM \\\hline
		Without timbre descriptors & 0.0432 &  0.9515\\
		With timbre descriptors & 0.0518 & 0.9414 \\ \hline
	\end{tabular}
	\label{tab:rec}
\end{table}

\section{Acknowledgments}
This work was funded by Science Foundation Ireland through the SFI Centre for Research Training in Machine Learning (18/CRT/6183)
\bibliographystyle{IEEEbib}
\bibliography{DAFx23_tmpl} 
\end{document}
